\documentclass[12pt]{article}
\newtheorem{thm}{Theorem}[section]

\newtheorem{rem}[thm]{Remark}

\def\baselinestretch{1.5}
\begin{document}

\begin{center}
{\Large Discretization of hyperbolic type Darboux integrable equations preserving integrability}
\end{center}

\begin{center}
{Ismagil Habibullin}\footnote{e-mail: habibullinismagil@gmail.com}

{Ufa Institute of Mathematics, Russian Academy of Science,\\
Chernyshevskii Str., 112, Ufa, 450077, Russia}\\

\bigskip

{Natalya Zheltukhina}\footnote{e-mail: natalya@fen.bilkent.edu.tr}

{Department of Mathematics, Faculty of Science,
 \\Bilkent University, 06800, Ankara, Turkey \\}

\bigskip

{Alfia Sakieva}\footnote{e-mail: alfiya85.85@mail.ru}

{Ufa Institute of Mathematics, Russian Academy of Science,\\
Chernyshevskii Str., 112, Ufa, 450077, Russia}\\

\end{center}

\begin{abstract}
A method of integrable discretization of the Liouville type nonlinear partial differential equations is suggested based on integrals. New examples of discrete Liouville type models are presented.
\end{abstract}

%{\it Keywords:} 
\def\baselinestretch{1.5}

\section{Introduction}

The problem of integrable discretization of the integrable PDE is very complicated and not enough studied. The same is true for evaluating the continuum limit for discrete models \cite{Kalyakin}.  In the present paper we undertake an attempt to clarify the connection between Liouville type partial differential equations and their discrete analogues. One unexpected observation is that there are pairs of equations, one continuous and the other one 
semi-discrete, having a common integral. Inspired by these examples, we  introduced a method of discretization of PDE having a nontrivial integral. Similar ideas are used in  \cite{Winternitz} where a method of construction of difference scheme for ordinary differential equations preserving the classical Lie group is suggested.  Let us begin with the necessary definitions.

We consider 
discrete equations of the form 
\begin{equation}\label{discrete}
v(n+1,m+1)=f(v(n,m),v(n+1,m),v(n,m+1))
\end{equation}
and 
semi-discrete 
chains
\begin{equation}\label{semi-discrete}
t(n+1,x)=f(x,t(n,x),t(n+1,x),t_x(n,x))\,.
\end{equation}
Equations (\ref{discrete}) and (\ref{semi-discrete}) are discrete and 
 semi-discrete analogues of  
hyperbolic equations 
\begin{equation}\label{hyperbolic}
u_{xy}=f(x,y,u,u_x,u_y)\, .
\end{equation}
Functions 
$v=v(n,m)$, 
$t=t(n,x)$ and $u=u(x,y)$  depend on 
discrete variables $n$
and $m$ 
and continuous variables $x$ and $y$. 
Through the paper we use  the following notations: 
$$v_{i,j}=v(n+i,m+j); \qquad v_i=v_{i,0}; \qquad \bar{v}_{j}=v_{0,j}; \qquad t_{i}=t(n+i,x)\, .$$

For equation (\ref{hyperbolic})  function $W(x,y,u,u_y, u_{yy}, \ldots, \partial^k u/\partial y^k)$ is called an $x$-integral of order $k$
if $D_xW=0$ and $W_{\partial^k u/\partial y^k}\ne 0$, and 
 function $\bar{W}(x,y,u, u_x, u_{xx},\ldots,  \partial^m u/\partial x^m)$ is called a $y$-integral of order $m$ if $D_y\bar{W}=0$ and $\bar{W}_{\partial^m u/\partial x^m}\ne 0$.
Here, $D_x$ and $D_y$ denote the total derivatives with respect to $x$ and $y$. Equation (\ref{hyperbolic}) is called Darboux integrable if it possesses  nontrivial $x$- and $y$- integrals.

For equation (\ref{semi-discrete})  function $F(x,n,t_m, t_{m+1}, t_{m+2}, \ldots, t_{m'})$ is called an $x$-integral of order
$m'-m+1$  if $D_xF=0$ and $F_{t_m}\ne 0$, $F_{t_m'}\ne 0$  and function $I(x, n, t, t_x, t_{xx}, \ldots, \frac{d^k t}{dt^k})$ is called an $n$-integral of order $k$, if $DI=I$ and $I_{\frac{d^k t}{dt^k}}\ne 0$. 
Here, $D$ is the forward shift operator in $n$, i.e. $Dh(n, x)=h(n+1, x)$.   Equation (\ref{semi-discrete}) is called Darboux integrable if it possesses  nontrivial $x$- and $n$- integrals. 

For equation (\ref{discrete}) function $I(n,m, \bar v_{k}, \bar{v}_{k+1}, \ldots, \bar{v}_{k'})$ is called an $n$-integral of order $k'-k+1$ if
$DI=I$ and $I_{\bar{v}_{k}}\ne 0$, $I_{\bar{v}_{k'}}\ne 0$, and function $\bar{I}(n, m, v_{r}, v_{r+1}, v_{r+2}, \ldots, v_{r'})$ is called an $m$-integral of order $r'-r+1$, if $\bar{D}\bar{I}=\bar{I}$ and $\bar{I}_{v_r}\ne 0$, $\bar{I}_{v_{r'}}\ne 0$. 
Here,  $D$ and $\bar{D}$ are the forward shift operators in $n$ and $m$ respectively.  Equation (\ref{discrete}) is called Darboux integrable, if it possesses nontrivial $n$- and $m$-integrals (see also \cite{AdlerStartsev}).

Continuous equations  (\ref{hyperbolic}) are very-well studied. In particular, the question of describing all
Darboux integrable equations  (\ref{hyperbolic})  is completely solved ( see \cite{{GOURSAT}} - \cite{Zhiber1}).
All equations  (\ref{hyperbolic}) possessing $x$- and $y$-integrals of order 2 are described 
by the following theorem. 
\begin{thm} \label{ThZh}(see \cite{Zhiber1}) Any equation (\ref{hyperbolic}), for which there exist 
second order $x$- and $y$-integrals, under the change of variables 
$x\to X(x)$, $y\to Y(y)$, $u\to U(x,y, u)$, can be reduced to one of the kind:\\
(1) $u_{xy}=e^u$, $\bar{W}=u_{xx}-0.5u_x^2$,  $W=u_{yy}-0.5u_y^2$;\\
(2) $u_{xy}=e^yu_y$, $\bar{W}=u_x-e^u$, $W=\frac{u_{yy}}{u_y}-u_y$;\\
(3) $u_{xy}=e^u\sqrt{u_y^2-4}$, $\bar{W}=u_{xx}-0.5u_x^2-0.5e^{2u}$, $W=\frac{u_{yy}-u_y^2+4}{\sqrt{u_y^2-4}}$;\\
(4) $u_{xy}=u_xu_y\left(\frac{1}{u-x}-\frac{1}{u-y}\right)$, $\bar{W}=\frac{u_{xx}}{u_x}-\frac{2u_x}{u-x}+\frac{1}{u-x}$, $W=\frac{u_{yy}}{u_y}-\frac{2u_y}{u-y}+\frac{1}{u-y}$;\\
(5) $u_{xy}=\psi(u)\beta(u_x)\bar{\beta}(u_y)$, $(ln\psi)''=\psi^2$, $\beta\beta'=-u_x$,  
$\bar{\beta}\bar{\beta}'=-u_y$, \\
 $\qquad \qquad \bar{W}=\frac{u_{xx}}{\beta(u_x)}-\psi(u)\beta(u_x)$, $W
=\frac{u_{yy}}{\bar{\beta}(u_y)}-\psi(u)\bar{\beta}(u_y)$;\\
(6) $u_{xy}=\frac{\beta(u_x)\bar{\beta}(u_y)}{u}$, $\beta\beta'+c\beta=-u_x$, $\bar{\beta}\bar{\beta}'+c\bar{\beta}=-u_y$,\\
$ \qquad \qquad \bar{W}=\frac{u_{xx}}{\beta}-\frac{\beta}{u}$, $W=\frac{u_{yy}}{\bar{\beta}}-\frac{\bar{\beta}}{u}$;\\
(7) $u_{xy}=-2\frac{\sqrt{u_xu_y}}{x+y}$, $\bar{W}=\frac{u_{xx}}{\sqrt{u_x}}+2\frac{\sqrt{u_x}}{x+y}$,  
$W=\frac{u_{yy}}{\sqrt{u_y}}+2\frac{\sqrt{u_y}}{x+y}$;\\
(8) $u_{xy}=\frac{1}{(x+y)\beta(u_x)\bar{\beta}(u_y)}$, $\beta'=\beta^3+\beta^2$, 
$\bar{\beta}'=\bar{\beta}^3+\bar{\beta}^2$,\\
$\qquad \qquad \bar{W}=u_{xx}\beta(u_x)-\frac{1}{(x+y)\beta(u_x)}$,
$W=u_{yy}\bar{\beta}(u_y)-\frac{1}{(x+y)\bar{\beta}(u_y)}$.
\end{thm}

On the contrary, the problem of describing all equations (\ref{discrete}) or (\ref{semi-discrete}) possessing both integrals    (so-called Darboux integrable equations) is very far from being solved (the problem of classification is solved only for a very special kind of semi-discrete equations \cite{HZhP}), it would be  beneficial for further classification to obtain new Darboux-integrable equations (\ref{discrete}) and semi-discrete chains
(\ref{semi-discrete}).  
It was observed that many chains (\ref{semi-discrete}) and their continuum limit equations (\ref{hyperbolic})
possess the same  $n$- and $y$-integrals: 
$$
\begin{array}{|l|l|l|l|}
\hline
semi-discrete\, \,chain&n-integral\, I&continuous  &y-integral \,\bar{W}\\
&&analogue&\\
\hline
t_{1x}=t_x+0.5t_1^2-0.5t^2&t_x-0.5t^2&u_{xy}=uu_y& u_x-0.5u^2\\
\hline
t_{1x}=t_x+Ce^{0.5(t+t_1)}, C=Const&t_{xx}-0.5t_x^2&u_{xy}=e^u&u_{xx}-0.5u_x^2\\
\hline
t_{1x}=t_x+\sqrt{e^{2t}+Ce^{t+t_1}+e^{2t_1}}&t_{xx}-0.5t_x^2-0.5e^{2t}&u_{xy}=e^u\sqrt{1+u_y^2}&u_{xx}-0.5u_x^2-0.5e^{2u}\\
\hline
\end{array}
$$

The main aim of the present paper is the discretization of equations (\ref{hyperbolic}) preserving the structure of $y$-integrals of order 2: we take $y$-integral for each of eight classes of Theorem \ref{ThZh} and find the semi-discrete chain (\ref{semi-discrete}) possessing the given $n$-integral ($y$-integral).
The next Theorem presents a list of semi-discrete models of Darboux integrable equations (\ref{hyperbolic}) from 
Theorem \ref{ThZh}  with  integrals of order 2. 
\begin{thm} \label{ThSemi1} Below is the list of  
equations (\ref{semi-discrete}) possessing the given $n$-integral $I$ :
$$
\begin{array}{|l|l|l|}
\hline
given \, \, n-integral&the \,\, corresponding \, \,chain &\\
\hline
I=t_{xx}-0.5t_x^2& t_{1x}=t_x+Ce^{0.5(t+t_1)}, \, C=Const&(1^*)\\
\hline
I=t_x-e^t&t_{1x}=t_x-e^t+e^{t_1}&(2^*a)\\
\hline
I=\frac{t_{xx}}{t_x}-t_x&t_{1x}=K(t, t_1)t_x, \, where\, \,K_tK^{-1}+K_{t_1}=K-1&(2^*b)\\
\hline
I=t_{xx}-0.5t_x^2-0.5e^{2t}&t_{1x}=t_x+\sqrt{e^{2t}+Re^{t+t_1}+e^{2t_1}}, \, R=Const&(3^*a)\\
\hline
 I=\frac{t_{xx}-t_x^2+4}{\sqrt{t_x^2-4}}&t_{1x}=(1+Re^{t+t_1})t_x+\sqrt{R^2e^{2(t+t_1)}+2Re^{t+t_1}}
\sqrt{t_x^2-4} &(3^*b)\\
\hline
I=\frac{t_{xx}}{t_x}-\frac{2t_x}{t-x}+\frac{1}{t-x}&t_{1x}=\frac{(t_1+L)(t_1-x)}{(t+L)(t-x)}t_x, \, L=Const&(4^*)\\
\hline
I=\frac{t_{xx}}{\beta(t_x)}-\psi(t){\beta(t_x)}&   \beta(t_x)=it_x \,\, and \, \,t_{1x}=K(t,t_1)t_x,\,\, where &(5^*)
\\ 
(ln\psi)''=\psi^2, 
\beta\beta'=-t_x& K_t+KK_{t_1}+K^2\psi(t_1)-K\psi(t)=0 \, &
\\
\hline
 I=\frac{t_{xx}}{\beta(t_x)}-\frac{\beta(t_x)}{t}, & \beta(t_x)=Rt_x,\, \, and \, \,t_{1x}=K(t,t_1)t_x,\,\, where &(6^*)\\
\beta\beta'+c\beta=-t_x&\frac{K_t}{K}+K_{t_1}=\frac{R^2(tK-t_1)}{tt_1}&\\
\hline
I=\frac{t_{xx}}{\sqrt{t_x}}+\frac{2\sqrt{t_x}}{x+R}& t_{1x}=\left(\sqrt{t_x}+\frac{C}{x+R}\right)^2, \, R=Const,\, C=Const&(7^*)\\
\hline
 I=\beta(t_x)t_{xx}-\frac{1}{(x+R)\beta(t_x)}, &\beta(t_x)=-1 \, \,and \,\, t_{1x}=t_x+\frac{t_1-t+C}{x+R}&(8^*)\\
\beta'=\beta^3+\beta^2&&\\
\hline
\end{array}
$$
\end{thm}
It is remarkable that each equation in Theorem \ref{ThSemi1} also admits a nontrivial $x$-integral. 
It means that discretization preserving the structure of $y$-integrals sends Darboux integrable 
equations (\ref{hyperbolic}) into Darboux integrable chains
(\ref{semi-discrete}). 

Note that equation ($1^{*}$) was found in \cite{BD}. Equation ($3^{*}a$) for $R=2$ was found 
in \cite{AdlerStartsev}, equations ($2^{*}a$) and ($3^{*}a$) are found in \cite{HZhP}. 
To our knowledge, the other equations from  Theorem \ref{ThSemi1} are new.

The next theorem lists $x$-integrals for chains from Theorem \ref{ThSemi1}. 

\begin{thm}\label{ThSemi2}

\noindent(I) The equations $(2^*b)$, $(5^*)$ and $(6^*)$ from Theorem \ref{ThSemi1} having the 
form $t_{1x}=K(t, t_1)t_x$ admit $x$-integral  $F(t,t_1)$, where function $F$ is a solution of  
$F_t+K(t,t_1)F_{t_1}=0$ with a given function $K(t,t_1)$.  \\
(II) $x$-integrals of equations  $(8^*)$, $(1^*)$,  $(3^*a)$, $(3^*b)$, $(4^*)$,  $(7^*)$ and  $(2^*a)$  
 are $F=(t_1-t+C)/(x+y)$,
$F=e^{(t_1-t)/2}+e^{(t_1-t_2)/2}$, 
$F= arcsinh(ae^{t_1-t_2}+b)+arcsinh(ae^{t_1-t}+b)$ with $a=2(4-R^2)^{-1/2}$, $b=R(4-R^2)^{-1/2}$,
$F=\sqrt{Re^{2t_1}+2e^{t_1-t}}+\sqrt{Re^{2t_1}+2e^{t_1-t_2}}$, 
$F=(t_1-t)(t_2+L)(t_2-t)^{-1}(t_1+L)^{-1}$, 
 $F=(2t_1-t-t_2)/(2C^2)-1/(x+R)$  and $F=(e^{t}-e^{t_2})(e^{t_1}-e^{t_3})(e^{t}-e^{t_3})^{-1}(e^{t_1}-e^{t_2})^{-1}$ correspondingly.
\end{thm}

One can also apply the discretization method preserving the structure of integrals for semi-discrete chains 
(\ref{semi-discrete}): take $x$-integral for a semi-discrete chain and find discrete equation (\ref{discrete})
with the given $m$-integral ($x$-integral).

In spite of the absence of the complete classification for Darboux-integrable semi-discrete 
chains (\ref{semi-discrete}) there is a large variety of such chains in  literature 
(see, for instance, \cite{AdlerStartsev}, \cite{HZhP} and \cite{Startsev}). The procedure 
of obtaining fully discrete equations for a given integral is a difficult task and requires 
further investigation. As a rule it is reduced to a very complicated functional equation. 
We illustrate  the application of the discretization method on   chains ($1^{*}$), ($4^{*}$) and ($7^{*}$) 
from Theorem \ref{ThSemi1}.  The discrete analogues of the chains are presented in the next Remark.

\begin{rem}\label{Remark1} Below is the list of  
equations (\ref{discrete}) possessing the given $m$-integral $\bar{I}$ :
$$
\begin{array}{|l|l|l|}
\hline
given \, \, m-integral&the \,\, corresponding \,\, equation &\\
\hline
\bar{I}=e^{(v_1-v)/2}+e^{(v_1-v_2)/2} &e^{v_{1,1}+v}=\frac{1}{C+e^{-(v_1+\bar{v}_1)}} &(1^{**})\\
\hline
\bar{I}=(v_1-v)(v_2+L)(v_2-v)^{-1}(v_1+L)^{-1}& 
v_{1,1}=\frac{L(v_1+\bar{v}_1-v)+v_1\bar{v}_1}{L+v}
&
(4^{**})\\
\hline
\bar{I}=2v_1-v-v_2&v_{1,1}=v_1+h(\bar{v}_1-v), \, 
z=h(2z-h(z))&(7^{**})\\
\hline
\end{array}
$$
The equations $(1^{**})$,  $(4^{**})$ and   $(7^{**})$ have respectively 
the following $n$-integrals
$I= e^{(\bar{v}_1-v)/2}+e^{(\bar{v}_1-\bar{v}_2)/2}$,
$I=(\bar{v}_1-\bar{v})(\bar{v}_2+L)(\bar{v}_2-t)^{-1}(\bar{v}_1+L)^{-1}$ and
$I= \bar{v}_1-v-h^{-1}(\bar{v}_1-v)$ with $h^{-1}$ being the inverse function of function $h$ 
that satisfies the  
functional equation $z=h(2z-h(z))$.
\end{rem}

Equation ($1^{**}$) from Remark \ref{Remark1} appeared in \cite{hirota}, 
equations ($4^{**}$) and ($7^{**}$) seem to be new, unfortunately we failed to answer the question whether equation $z=h(2z-h(z))$ has any solution different from linear one $h(z)=z+C$.

The article is organized as follows. Theorem \ref{ThSemi1} is proved in Section 2. 
The proof of Theorem \ref{ThSemi2} is omitted.  Chains ($1^*$),
($2^*a$) and ($3^*a$) are of the form $t_{1x}=t_x+d(t,t_1)$, and their 
$x$-integrals can be seen in \cite{HZhP}. One can  find $x$-integrals for  chains ($3^*b$), 
($4^*$), ($7^*$) and ($8^*$)
 by direct calculations. In Section 3 the discretization of chains ($1^*$),
($4^*$) and ($7^*$)  from Remark \ref{Remark1} are presented and for each obtained discrete  equation the second 
integral is found. In Section 4 the Conclusion is drawn.

\section{Proof of Theorem \ref{ThSemi1}}
 
\noindent \underline{Case $(1^*)$}:
Consider all chains (\ref{semi-discrete}) with $n$-integral of the form
$I=t_{xx}-\frac{1}{2}{t_x}^2$. Equality $DI=I$ implies
\begin{equation}\label{(1)}f_x+f_tt_x+f_{t_1}f+f_{t_x}t_{xx}-\frac{1}{2}f^2=t_{xx}-\frac{1}{2}{t_x}^2. 
\end{equation}  
By comparing the coefficients before $t_{xx}$ in (\ref{(1)}) we have
$f_{t_x}=1$. Therefore, 
\begin{equation}\label{(2)}f(x,t,t_1,t_x)=t_x+d(x,t,t_1)\,.
\end{equation}
 We
substitute (\ref{(2)}) into (\ref{(1)}) and get
$d_x+d_tt_x+d_{t_1}t_x+d_{t_1}d-\frac{1}{2}{t_x}^2-dt_x-\frac{1}{2}d^2=-\frac{1}{2}{t_x}^2$,
or equivalently,
$d_t+d_{t_1}-d=0$ 
  and $d_x+d_{t_1}d-\frac{1}{2}d^2=0$.
   We solve the last two equations simultaneously and find that   $d={e^{t_1}}K(x,t_1-t)$, where 
   $K=Ce^{-\frac{1}{2}(t_1-t)}$ and $C$ is an arbitrary
   constant. Therefore, chain (\ref{semi-discrete}) with
   $n$-integral $I=t_{xx}-\frac{1}{2}{t_x}^2$ becomes
   $t_{1x}=t_x+Ce^{(t_1+t)/2}$. 
   
\noindent \underline{Case $(2^*a)$}:
 Consider all chains (\ref{semi-discrete}) with $n$-integral 
$I=t_x-e^{t}$.
Equality $DI=I$
implies 
$f-e^{t_1}=t_x-e^{t}$, which gives the equation
$t_{1x}=f=t_x-e^{t}+e^{t_1}$.

\noindent \underline{Case $(2^*b)$}:
Consider all chains (\ref{semi-discrete}) with
$n$-integral $I=\frac{t_{xx}}{t_x}-t_x$.
Equality
$DI=I$ implies
\begin{equation}\label{(5)}\frac{f_x+f_tt_x+f_{t_1}f+f_{t_x}t_{xx}}{f}-f=\frac{t_{xx}}{t_x}-t_x.
\end{equation}
By comparing the coefficients before $t_{xx}$ in (\ref{(5)}) we have
$f_{t_x}/f=1/t_x$, that is 
$f=K(x,t,t_1)t_x$.
Substitute $f=K(x,t,t_1)t_x$ into (\ref{(5)}) and have
$\frac{K_x}{K}+{\frac{K_t}{K}}t_x+K_{t_1}t_x-Kt_x=-t_x$, 
or
equivalently (by comparing the coefficients before $t_x$ and
${t_x}^0$), we get 
$\frac{K_t}{K}+K_{t_1}=K-1$ and $K_x=0$.
Therefore,  equations $t_{1x}=K(t,t_1)t_x$, where K satisfies
$\frac{K_t}{K}+K_{t_1}=K-1$
are the only chains (\ref{semi-discrete}) that admit  $n$-integral $I$ of the form $I=\frac{t_{xx}}{t_x}-t_x$.

\noindent \underline{Case $(3^*a)$}:
Consider all chains (\ref{semi-discrete}) with
$n$-integral $I=t_{xx}-\frac{1}{2}{t_x}^2-\frac{1}{2}e^{2t}$. 
Equality 
$DI=I$ implies
\begin{equation}\label{(7)}
f_x+f_tt_x+f_{t_1}f+f_{t_x}t_{xx}-\frac{1}{2}f^2-\frac{1}{2}e^{2t_1}=t_{xx}-\frac{1}{2}{t_x}^2-\frac{1}{2}e^{2t}.
\end{equation}
By comparing the coefficients before $t_{xx}$ in (\ref{(7)}) we have 
$f_{t_x}=1$, that is 
$f(x,t,t_1,t_x)=t_x+d(x,t,t_1)$. 
Substitute  $f(x,t,t_1,t_x)=t_x+d(x,t,t_1)$ into (\ref{(7)}) and have
\begin{equation}\label{(10)}
d_x+d_tt_x+d_{t_1}(t_x+d)-\frac{1}{2}(t_x+d)^2-\frac{1}{2}e^{2t_1}=-\frac{1}{2}{t_x}^2-\frac{1}{2}e^{2t}.
\end{equation}
Compare the coefficients before $t_x$ and ${t_x}^0$ in (\ref{(10)}) and get
\begin{equation}
\label{(11)}
d_t+d_{t_1}-d=0, \qquad 
d_x+d_{t_1}d-\frac{1}{2}d^2-\frac{1}{2}e^{2t_1}=-\frac{1}{2}e^{2t}. 
\end{equation}
The first equation in  (\ref{(11)}) has a solution $d=e^{t_1}K(x,t_1-t)$. 
Substitution of   this expression into the second equation of (\ref{(11)})
gives 
$e^{-t_1}K_x+K_{t_1-t}K+\frac{1}{2}K^2-\frac{1}{2}+\frac{1}{2}e^{-2(t_1-t)}=0$.
Since $K$ depends on $U=t_1-t$ and $x$, then $K_x=0$ and the last equation becomes 
$2K'K+K^2=1-e^{-2U}$, and hence, 
$d=e^{t_1}K=\sqrt{e^{2t_1}+e^{2t}+Re^{t+t_1}}$, where $R$ is and arbitrary constant. 
Therefore, chain (\ref{semi-discrete})  with $n$-integral 
$I=t_{xx}-\frac{1}{2}{t_x}^2-\frac{1}{2}e^{2t}$ becomes 
$t_{1x}=t_x+\sqrt{e^{2t_1}+e^{2t}+Re^{t+t_1}}, R=const$.

\noindent \underline{Case $(3^*b)$}:
 Consider all chains (\ref{semi-discrete}) with
$n$-integral $I=\frac{t_{xx}-{t_x}^2+4}{\sqrt{{t_x}^2-4}}$.
Equality
$DI=I$ implies
\begin{equation}\label{(13)}
\frac{f_x+f_tt_x+f_{t_1}f+f_{t_x}t_{xx}-f^2+4}{\sqrt{f^2-4}}=\frac{t_{xx}-{t_x}^2+4}{\sqrt{{t_x}^2-4}}.
\end{equation}
By comparing the coefficients before $t_{xx}$ in (\ref{(13)}) we get 
$\frac{f_{t_x}}{\sqrt{f^2-4}}=\frac{1}{\sqrt{{t_x}^2-4}}$, that is
$arccosh{\frac{f}{2}}=arccosh\frac{t_x}{2}+K(x,t,t_1)$.
Thus, 
\begin{equation}\label{(14)}
f(x,t,t_1,t_x)=At_x+B\sqrt{{t_x}^2-4},
\end{equation}
where $A(x,t,t_1)=cosh K$, $B(x,t,t_1)=sinhK$, $A^2-B^2=1$.
 Note that $f=2 cosh((arccosh\frac{t_x}{2})+K)$,
i.e. $\sqrt{f^2-4}=2sinh((arccosh\frac{t_x}{2})+K)=2(\sqrt{\frac{{t_x}^2}{4}-1}coshK+\frac{t_x}{2}sinhK)$, or $\sqrt{f^2-4}=Bt_x+A\sqrt{{t_x}^2-4}$.
Substitute (\ref{(14)}) into (\ref{(13)}) and have
$$t_xA_x+B_x\sqrt{{t_x}^2-4}+{t_x}^2A_t+t_xB_t\sqrt{{t_x}^2-4}+
(t_xA_{t_1}+B_{t_1}\sqrt{{t_x}^2-4})(At_x+B\sqrt{{t_x}^2-4})
$$
$$-(At_x+B\sqrt{{t_x}^2-4})^2+4=-(Bt_x+A\sqrt{{t_x}^2-4})\sqrt{{t_x}^2-4},$$ 
that can be written shortly as
\begin{equation}\label{equation}
({t_x}^2-4)(\alpha_{1}+\alpha_{2}t_x)^2=(\alpha_{3}+\alpha_{4}t_x+\alpha_{5}{t_x}^2)^2,
\end{equation}
where
$\alpha_1=B_x, \alpha_2=B_t+A_{t_1}B+B_{t_1}A-2AB+B, \alpha_{3}=-4B_{t_1}B+4B^2+4-4A, \alpha_{4}=A_x, \alpha_{5}=A_t+A_{t_1}A+B_{t_1}B-A^2-B^2+A$.
We compare the coefficients before ${t_x}^4$, ${t_x}^3$, ${t_x}^2$, $t_x$, $t_x^0 $ in  (\ref{equation}) and have ${\alpha_{2}}^2={\alpha_{5}}^2$, $2{\alpha_{1}}{\alpha_{2}}=2{\alpha_{4}}{\alpha_{5}}$,  ${\alpha_{1}}^2-4{\alpha_{2}}^2={\alpha_4}^2+2{\alpha_{3}}{\alpha_{5}}$,  $-8{\alpha_{1}}{\alpha_{2}}=2{\alpha_{3}}{\alpha_{4}}$, 
$-4{\alpha_{1}}^2={\alpha_{3}}^2$, that implies $\alpha_{1}=\alpha_{2}=\alpha_{3}=\alpha_{4}=\alpha_{5}=0$,
which is possible only if  $A=1+Re^{t+t_1}$ and $B=\sqrt{R^2e^{2(t+t_1)}+2R {e^{(t+t_1)}}}$, where $R=const$.
Therefore, by (\ref{(14)}), the chain  (\ref{semi-discrete}) with $n$-integral  
$I=\frac{t_{xx}-{t_x}^2+4}{\sqrt{{t_x}^2-4}}$
becomes $t_{1x}=(1+Re^{t+t_1})t_x+\sqrt{R^2e^{2(t+t_1)}+2R {e^{(t+t_1)}}}\sqrt{{t_x}^2-4}$. 

\noindent \underline{Case $(4^*)$}:
Consider chains (\ref{semi-discrete}) with
$n$-integral $I=\frac{t_{xx}}{t_x}-\frac{2t_x}{t-x}+\frac{1}{t-x}$.
Equality
$DI=I$ implies
\begin{equation}\label{(15)}
\frac{f_x+f_tt_x+f_{t_1}f+f_{t_x}t_{xx}}{f}-
\frac{2f}{t_1-x}+\frac{1}{t_1-x}=\frac{t_{xx}}{t_x}-\frac{2t_x}{t-x}+\frac{1}{t-x}.
\end{equation}
We compare the coefficients before $t_{xx}$ and have
$f_{t_x}/f=1/t_x$, that is 
$f=t_xK(x,t,t_1)$. 
Substitute $f=t_xK$ into (\ref{(15)}) and have
\begin{equation}\label{(17)}
\frac{K_xt_x+K_t{t_x}^2+K_{t_1}K{t_x}^2}{Kt_x}-\frac{2Kt_x}{t_1-x}+
\frac{1}{t_1-x}=-\frac{2t_x}{t-x}+\frac{1}{t-x}.
\end{equation}
By comparing the coefficients before $t_x$ and ${t_x}^0$ in (\ref{(17)}) we get
\begin{equation}\label{(18)}
\frac{K_t}{K}+K_{t_1}=\frac{2K}{t_1-x}-\frac{2}{t-x}, \qquad
\frac{K_x}{K}=-\frac{1}{t_1-x}+\frac{1}{t-x}.\end{equation}
We solve two equations of (\ref{(18)}) simultaneously  and have 
$K=\frac{t_1+L}{t+L}\frac{t_1-x}{t-x}$, where $L$ is an arbitrary constant. 
Therefore, any chain (\ref{semi-discrete}) with $n$-integral 
$I=\frac{t_{xx}}{t_x}-\frac{2t_x}{t-x}+\frac{1}{t-x}$
 becomes $t_{1x}=\frac{t_1+L}{t+L}\frac{t_1-x}{t-x}t_x$.

\noindent \underline{Case $(5^*)$  }:
Consider all chains (\ref{semi-discrete}) with
$n$-integral $I=\frac{t_{xx}}{\beta}-\psi{\beta}$, where $\beta=\beta(t_x),\psi=\psi(t), \beta{\beta}'=-t_x$.
% and $(\ln \psi)''=\psi^2$.
We have,  $2\beta{\beta}'=-2{t_x}$, i.e. ${\beta}^2=-{t_x}^2+M^2$, or $\beta=\sqrt{M^2-{t_x}^2}$, where $M$
is an arbitrary constant.
Equality 
$DI=I$ implies
\begin{equation}\label{(25a)} \frac{f_x+f_t{t_x}+f_{t_1}f+f_{t_x}t_{xx}}{\beta(f)}-\psi(t_1)\beta(f)=\frac{t_{xx}}{\beta(t_x)}-
\psi(t)\beta(t_x).
\end{equation}
We compare the coefficients before $t_{xx}$ and have
$f_{t_x}/\beta(f)=1/\beta(t_x)$ which implies that\\ 
\underline{either 
$(5^*a)$}: $M=0$, $\beta(t_x)=it_x$ and $t_{1x}=K(x,t,t_1)t_x$,\\
\underline{ or 
 $(5^*b)$}: $M\ne 0$ and then 
$arcsin{\frac{f}{M}}=arcsin{\frac{t_x}{M}}+L(x,t,t_1)$, that is, 
\begin{equation}\label{5b}f=t_xA(x,t,t_1)+\sqrt{M^2-{t_x}^2}B(x,t,t_1), A^2+B^2=1.\end{equation}
In case $(5^*a)$ we substitute  $t_{1x}=f=K(x,t,t_1)t_x$ into (\ref{(25a)}), use that $\beta(t_x)=it_x$, and obtain
\begin{equation}\label{5aK} K_x=0, \qquad \frac{K_t}{K}+K_{t_1}+\psi(t_1)K=\psi(t).\end{equation}
Therefore, the chains (\ref{semi-discrete}) with $n$-integral $I=\frac{t_{xx}}{it_x}-i\psi(t)t_x$ 
are  equations $t_{1x}=K(t,t_1)t_x$, where function $K$ satisfies (\ref{5aK}).\\
Let us consider case $(5^*b)$.
 Note that 
 $$M^2-f^2=M^2-A^2{t_x}^2-2ABt_x\sqrt{M^2-{t_x}^2}-B^2M^2+B^2{t_x}^2=(Bt_x-A\sqrt{M^2-{t_x}^2})^2$$
and $\beta(f)=\pm(Bt_x-A\sqrt{M^2-{t_x}^2})$, $\beta(t_x)=\sqrt{M^2-{t_x}^2}$.
Substitute (\ref{5b}) into (\ref{(25a)}) and get
$$\frac{A_x{t_x}+B_x\sqrt{M^2-{t_x}^2}+A_t{t_x}^2+B_tt_x\sqrt{M^2-{t_x}^2}+(A_{t_1}t_x+
B_{t_1}\sqrt{M^2-{t_x}^2})(At_x+B\sqrt{M^2-{t_x}^2})}{\pm(Bt_x-A\sqrt{M^2-{t_x}^2})}
$$
$$=
\pm(Bt_x-A\sqrt{M^2-{t_x}^2})\psi(t_1)-\sqrt{M^2-{t_x}^2}\psi(t),$$
or the same,
\begin{equation}\label{5bshort}(M^2-{t_x}^2)({\alpha}_1+{\alpha}_2{t_x})^2=({\alpha}_3+{\alpha}_4{t_x}+
{\alpha}_5{t_x}^2)^2,\end{equation}
where ${\alpha}_1=B_x$, ${\alpha}_4=A_x$, ${\alpha}_2=B_t+A_{t_1}B+AB_{t_1}+2AB\psi(t_1)+B\psi(t)$, ${\alpha}_3=BB_{t_1}M^2-A^2M^2\psi(t_1)-A\psi(t)M^2$, 
${\alpha}_5=A_t+A_{t_1}A-B_{t_1}B-B^2\psi(t_1)+A^2\psi(t_1)+A\psi(t)$.
We compare the coefficients before $t_{x}^k$, $k=0,1,2,3,4$, in (\ref{5bshort}) and find that 
${\alpha}_1={\alpha}_2={\alpha}_3={\alpha}_4={\alpha}_5=0$,which is possible only if $\psi=R$ is a constant function, that contradicts 
to the equation $(ln\psi)''=\psi^2$. Therefore, case 
($5^*b$) is not realized.   
   
\noindent \underline{Case $(6^*)$}: Consider chains (\ref{semi-discrete}) with
$n$-integrals $I=\frac{t_{xx}}{\beta(t_x)}-\frac{\beta(t_x)}{t}$, where $\beta=\beta(t_x)$ and
$\beta\beta'+c\beta =-t_x$. The equality $DI=I$ implies 
\begin{equation}\label{6**} \frac{f_{t_x}}{\beta(f)}=\frac{1}{\beta(t_x)}\end{equation}
and 
\begin{equation}\label{6*}
\frac{f_x+f_t t_x+f_{t_1}f}{\beta(f)}-\frac{\beta(f)}{t_1}=-\frac{\beta(t_x)}{t}.
\end{equation}
Differentiation of (\ref{6**}) with respect to $x$, $t$, $t_1$ gives
\begin{equation}\label{6many}
f_{xt_x}=\frac{\beta'(f)}{\beta(t_x)}f_x, \qquad f_{t_xt}=\frac{\beta'(f)}{\beta(t_x)}f_t, \qquad
f_{t_xt_1}=\frac{\beta'(f)}{\beta(t_x)}f_{t_1}.
\end{equation}
First we differentiate (\ref{6*}) with  respect to $t_x$, use (\ref{6many}), and get
\begin{equation}\label{6*'}
\frac{1}{\beta(f)}f_t+\frac{1}{\beta(t_x)}f_{t_1}=-\frac{(c\beta(f)+f)}{t_1\beta(t_x)}+\frac{c}{t}+
\frac{t_x}{t\beta(t_x)}.
\end{equation}
Next we differentiate (\ref{6*'}) with respect to $t_x$, use (\ref{6many}), and arrive to the equality
$$\left\{\frac{t_x}{\beta(t_x)}-\frac{f}{\beta(f)} \right\}f_{t_1}=-\frac{(c\beta(f)+f)t_x}{t_1\beta(t_x)}
-\frac{\beta(f)}{t_1}+\frac{\beta(t_x)}{t}+\frac{ct_x}{t}+\frac{t_x^2}{t\beta(t_x)}.$$
There are two possibilities: \\
\underline{either ($6^*a$)}, when 
\begin{equation}\label{6a}
A:=\frac{t_x}{\beta(t_x)}-\frac{f}{\beta(f)} =0,
\end{equation} 
\underline{or ($6^*b$)}, when
\begin{equation}\label{6A}
f_{t_1}=\frac{\beta(f)\beta(t_x)}{t_x\beta(f)-f\beta(t_x)}
\left\{
\frac{-(c\beta(f)+f)t_x}{t_1\beta(t_x)}
-\frac{\beta(f)}{t_1}+\frac{\beta(t_x)}{t}+\frac{ct_x}{t}+\frac{t_x^2}{t\beta(t_x)}
\right\}.
\end{equation}
Let us first consider case ($6^*a$). 
It follows 
from (\ref{6a}) and (\ref{6**}) that
%from (\ref{6A}) and (\ref{6*'}) that 
$f_{t_x}/f=1/t_x$, that is $f=K(x,t,t_1)t_x$. We substitute $f=K(x,t,t_1)t_x$ into 
(\ref{6*}), use $\beta(f)/t_1=(\beta(t_x)f)/(t_xt_1)=\frac{\beta(t_x)K}{t_1}$, and obtain
$$
K_x+t_x\left\{\frac{K_t}{K}+K_{t_1}\right\}=\frac{\beta^2(t_x)}{t_x}\left\{ 
\frac{K}{t_1}-\frac{1}{t}\right\}, 
$$
that is, $K_x=0$, $\beta(t_x)=\sqrt{R^2t_x^2+Ct_x}$, $R=Const$, $B=Const$, and 
\begin{equation}\label{E}
 \frac{K_t}{K}+K_{t_1}=R^2\left\{\frac{K}{t_1}-\frac{1}{t}\right\}.
\end{equation}
Substitution of $\beta(t_x)=\sqrt{R^2t_x^2+Ct_x}$ into (\ref{6a}) shows that $\beta(t_x)=Rt_x$.
Therefore, in case ($6^*a$), the $n$-integral is $I=\frac{t_{xx}}{Rt_x}-\frac{Rt_x}{t}$ and the corresponding
chain (\ref{semi-discrete}) is of the form $t_{1x}=K(t,t_1)t_x$, where $K$ satisfies
(\ref{E}).\\
Let us now study case ($6^*b$). It follows from
(\ref{6A}) and (\ref{6*'})
that
\begin{equation}\label{6B}
f_t=\frac{f\beta(t_x)\beta(f)}{\beta(f)t_x-f\beta(t_x)}\left\{
\frac{c\beta(f)+f}{t_1\beta(t_x)}-\frac{c}{t}-\frac{t_x}{t\beta(t_x)}+\frac{\beta^2(f)}{t_1f\beta(t_x)}-
\frac{\beta(f)}{tf}
\right\}.
\end{equation}
First we differentiate (\ref{6A}) with respect to $t$ and find $f_{t_1t}$, use the expression for
$f_t$ from (\ref{6B}) and $\beta'(f)=-(f+c\beta(f))/\beta(f)$ to express $f_{t_1t}$ 
in terms of $\beta(f)$, $\beta(t_x)$, $f$, $t$, $t_1$, $t_x$. Then 
we differentiate (\ref{6B}) with respect to $t_1$ and find $f_{tt_1}$, use the expression for
$f_{t_1}$ from (\ref{6A}) and $\beta'(f)=-(f+c\beta(f))/\beta(f)$ to express $f_{tt_1}$ 
in terms of $\beta(f)$, $\beta(t_x)$, $f$, $t$, $t_1$, $t_x$.\\
Direct calculations show that 
$$
f_{tt_1}-f_{t_1t}=\frac{2\beta(f)ct_x(\beta^2(f)+cf\beta(f)+f^2)(-tf+t_1(c\beta(t_x)+t_x))}{tt_1^2
(\beta(t_x)f-\beta(f)t_x)^2}.
$$
Equality $f_{tt_1}=f_{t_1t}$ yields 
(i) $\beta^2(f)+cf\beta(f)+f^2=0$, i.e. $\beta(f)=A f$, $\beta(t_x)=At_x$, where $A=\frac{-c\pm\sqrt{c^2-4}}{2}$, or
(ii) $f=t_1t^{-1}(c\beta(t_x)+t_x)$.

Let us consider case (i). It follows from (\ref{6**}) that $f=K(x,t,t_1)t_x$. The same considerations as in 
part ($6^*a$) show that the chain (\ref{semi-discrete}) in this case is $t_{1x}=K(t,t_1)t_x$, where function $K(t, t_1)$ satisfies
(\ref{E}).

Let us consider case (ii). It follows from (\ref{6**}) that $\beta(f)=t_1t^{-1}((1-c^2)\beta(t_x)-ct_x)$. We substitute this expression for $\beta(f)$ into (\ref{6*} ) and get
$c^2(2-c^2)\beta^2(t_x)+2c(1-c^2)t_x\beta(t_x)-c^2t_x^2=0$,
that implies that (I) $c=0$, (II) $c^2=2$, (III) $\beta(t_x)=\frac{c}{2-c^2}t_x$, or 
(IV) $\beta(t_x)=-\frac{1}{c}t_x$. Cases (II) and (IV) are not realized, each of them is incompatible with
$\beta\beta'+c\beta=-t_x$. Case (III) is realized only for $c=2$ (with $\beta(t_x)=-t_x$) and 
$c=-2$ (with $\beta(t_x)=t_x$).  Therefore, using $f=t_1t^{-1}(c\beta(t_x)+t_x)$ and the fact that 
$c=0$  (with $\beta(t_x)=\pm i t_x$) or $c=\pm 2(\beta(t_x)=-\pm t_x)$ we arrive to a chain (\ref{semi-discrete}) of the form
$t_{1x}=\pm \frac{t_1}{t}t_x$. Note that chains $t_{1x}=\pm t_1t^{-1}t_x$ with $\beta(t_x)=\pm t_x$ or $\beta(t_x)=\pm  i t_x$ is of the form $t_{1x}=K(t,t_1)t_x$, where $K$ satisfies (\ref{E}) with $R^2=1$ 
(for $t_{1x}=-t_1t^{-1}t_x$) or $R^2=-1$ 
(for $t_{1x}=t_1t^{-1}t_x$).

\noindent \underline{Case $(7^*)$}:
Consider chains (\ref{semi-discrete}) with
$n$-integral $I=\frac{t_{xx}}{\sqrt{t_x}}+2\frac{\sqrt{t_x}}{x+y}$, $y=Const$.
Equality
$DI=I$ implies 
\begin{equation}\label{(26)}
\frac{f_x+f_tt_x+f_{t_1}f+f_{t_x}t_{xx}}{\sqrt{f}}+
2\frac{\sqrt{f}}{x+y}=\frac{t_{xx}}{\sqrt{t_x}}+2\frac{\sqrt{t_x}}{x+y}.
\end{equation} 
By comparing the coefficients before $t_{xx}$ we have 
$f_{t_x}/\sqrt{f}={1}/{\sqrt{t_x}}$, or
\begin{equation}\label{(27)}
f=(\sqrt{t_x}+K(x,t,t_1))^2.
\end{equation}
Substitute (\ref{(27)}) into (\ref{(26)}) and get
$K_x+K_tt_x+K_{t_1}t_x+2K_{t_1}\sqrt{t_x}K+K_{t_1}K^2+\frac{K}{x+y}=0$.
We compare the coefficients before $\sqrt{t_x}$,  $t_x$, ${t_x}^0$ and have
$2K_{t_1}K=0$, i.e. $K=L(x,t)$;
$K_t+K_{t_1}=0$, i.e. $K=L(x)$; and
$K_x+K_{t_1}K^2+\frac{K}{x+y}=0$, i.e. $K=\frac{C}{x+y}$, $C=const$.
Therefore, chain (\ref{semi-discrete}) with $n$-integral $I=\frac{t_{xx}}{\sqrt{t_x}}+2\frac{\sqrt{t_x}}{x+y}$  
becomes $t_{1x}=(\sqrt{t_x}+\frac{C}{x+y})^2$, where $C$ and $y$ are arbitrary constants.
 
\noindent \underline{Case $(8^*)$}: Consider chains (\ref{semi-discrete}) with $n$-integral 
$I=\beta(t_x)t_{xx}-\frac{1}{(x+y)\beta(t_x)}$, where $y$  is an arbitrary constant 
and $\beta'(t_x)=\beta^3(t_x)+\beta^2(t_x)$. The equality $DI=I$ gives
$$
\beta(f)\{ f_x+f_t t_x+f_{t_1}f+f_{t_x}t_{xx}\}-\frac{1}{(x+y)\beta(f)}=\beta(t_x)t_{xx}-\frac{1}{(x+y)\beta(t_x)},
$$
 that implies
 \begin{equation}\label{8**}
 \beta(f)f_{t_x}=\beta(t_x)
 \end{equation}
and 
\begin{equation}\label{8*}
\beta(f)\{ f_x+f_t t_x+f_{t_1}f\}=\frac{1}{(x+y)\beta(f)}-\frac{1}{(x+y)\beta(t_x)}.
\end{equation}
Differentiate (\ref{8**}) with respect to $x$, $t$, $t_1$ and get
\begin{equation}\label{8many}
f_{x t_{x}}=-(\beta(f)+1)\beta(t_x)f_x, \quad f_{t t_{x}}=-(\beta(f)+1)\beta(t_x)f_t, \quad 
f_{ t_1 t_{x}}=-(\beta(f)+1)\beta(t_x)f_{t_1}.
\end{equation}
Now differentiate (\ref{8*}) with respect to $t_x$, we have
\begin{equation}\label{8*'}
\beta(f)f_t+\beta(t_x)f_{t_1}=\frac{1}{x+y}-\frac{\beta(t_x)}{(x+y)\beta(f)}.
\end{equation}
Differentiate (\ref{8*'}) with respect to $t_x$ and get 
$
f_{t_1}=-\frac{1}{(x+y)\beta(f)}.
$
The last equation together with (\ref{8*'}), (\ref{8**}) and (\ref{8*}) gives 
\begin{equation}\label{A1-3}
f_{t_1}=-\frac{1}{(x+y)\beta(f)}, \quad
f_t=\frac{1}{(x+y)\beta(f)}, \quad f_{t_x}=\frac{\beta(t_x)}{\beta(f)}
\end{equation}
and 
\begin{equation}\label{A4}
f_x=\frac{1}{x+y}\left\{\frac{1}{\beta^2(f)}-\frac{1}{\beta(f)\beta(t_x)}-\frac{t_x}{\beta(f)}+\frac{f}{\beta(f)}  \right\}.
\end{equation}
Since, by (\ref{A1-3}) and (\ref{A4}),   
$f_{t_1 x}-f_{x t_1}=\frac{1}{\beta^2(f)(x_y)^2}(\beta(f)+	1)$,
then 
$\beta(f)=-1$, and, therefore, by  (\ref{A1-3}), we have
$f_{t_1}=(x+y)^{-1}$, $f_{t}=-(x+y)^{-1}$, $f_{t_x}=1$. Hence, 
$f(x,t, t_1, t_x)=t_x+\frac{t_1-t}{x+y}+C(x)$. We substitute this expression for $f$ into (\ref{A4})
and obtain $C(x)=C(x+y)^{-1}$, where $C$ is an arbitrary constant. Therefore, with the $n$-integral
$I=\frac{t_{xx}}{\sqrt{t_x}}+2\frac{\sqrt{t_x}}{x+y}$ the chain (\ref{semi-discrete}) becomes 
$t_{1x}=t_x+\frac{t_1-t}{x+y}+C(x+y)^{-1}$, where $y$ is arbitrary constant.

\section{Proof of Remark \ref{Remark1}}

 \noindent\underline{Case $1^{**}$}:
 Consider all equations (\ref{discrete}) with $m$-integral  $\bar{I}=e^{v_1-v}+e^{v_1-v_2}$.
Denote by  $e^{-v_j}=w_j$, $j=0,1,2$, and  $e^{-\bar{v}_1}=\bar{w}_1$. In new variables  
$\bar{I}=\frac{v+v_2}{v_1}$ is an $m$-integral of
equation $w_{1,1}=g(w,w_1\bar{w}_1)$.
 $\bar{D}\bar{I}=\bar{I}$ implies 
 \begin{equation}\label{(28)}
\frac{w_2+w}{w_1}=\frac{g_1+\bar{w}_1}{g}.  
 \end{equation}
 We differentiate both sides of (\ref{(28)}) with respect to $w_2$ and apply the shift operator $D^{-1}$, we have
 $$
\frac{1}{w_1}=\frac{{g_1}_{w_2}}{g} \quad \Rightarrow \quad
 D^{-1}\left(\frac{1}{w_1}\right)=D^{-1}\left(\frac{{g_1}_{w_2}}{g}\right)\quad \Rightarrow \quad
g_{w_1}=\frac{\bar{w}_1}{w}.
$$
Therefore,
 \begin{equation}\label{(30)}
  g=\frac{\bar{w}_1w_1}{w}+c(w,\bar{w}_1), \qquad 
 g_1=\frac{gw_2}{w_1}+c(w_1,g).
\end{equation}
We substitute (\ref{(30)}) into (\ref{(28)}) and get 
   \begin{equation}\label{(31)}
   g\frac{w}{w_1}=c(w_1,g)+\bar{w}_1.
   \end{equation}
   Substitution of  (\ref{(30)}) into (\ref{(31)}) implies that 
   $c(w,\bar{w}_1)w=c(w_1,g)w_1$,
   or the same,
   $c(w,\bar{w_1})w=D(c(w,\bar{w}_1)w)$.
Suppose that equation $w_{1,1}=g(w,w_1\bar{w}_1)$ does not admit an $m$-integral of the first order, then $c(w,\bar{w}_1)w=D(c(w,\bar{w}_1)w)=C=const$. Thus,
 $c(w,\bar{w}_1)=C/w$. Finally, $g(w,w_1,\bar{w}_1)=\frac{\bar{w}_1w_1}{w}+Cw^{-1}$. Therefore, the equations 
(\ref{discrete}) with $m$-integral  $\bar{I}=e^{v_1-v}+e^{v_1-v_2}$ becomes 
$e^{v_{1,1}+v}=(C+e^{-(v_1+\bar{v}_1)})^{-1}$, where $C$ is an arbitrary constant.
Note that this equation is symmetric with respect to variables $v_1$ and $\bar{v}_1$. 
Therefore, $n$-integral for the equation can be obtained by simply changing in $m$-integral
variables $v_j$ into variables $\bar{v}_j$, $j=1,2$. 

\noindent \underline{Case $4^{**}$}:  Consider equations (\ref{discrete}) with $m$-integral 
$\bar{I}=\frac{(v_1-v)(v_2+L)}{(v_2-v)(v_1+L)}$. Equation $v_{1,1}=f(v,v_1,\bar{v}_1)$ can be rewritten as 
$v_{-1,1}=r(v, v_{-1},\bar{v}_1)$. Equality $\bar{D}\bar{I}=\bar{I}$ implies
\begin{equation}\label{4d1}
\frac{f-\bar{v}_1)(f_1+L)}{(f_1-\bar{v}_1)(f+L)}=\frac{(v_1-v)(v_2+L)}{(v_2-v)(v_1+L)}.
\end{equation}
Take  the logarithmic derivative of (\ref{4d1})  with respect to $v_2$ and then apply the shift operator $D^{-1}$, we get
\begin{equation}\label{4d2}
\frac{{f_{1}}_{v_2}}{f_1+L}-\frac{{f_{1}}_{v_2}}{f_1-\bar{v}_1}=\frac{1}{v_2+L}-\frac{1}{v_2-v} \quad
\Rightarrow \quad
\frac{f_{v_1}(r+L)}{(f+L)(f-r)}=\frac{v_{-1}+L}{(v_1+L)(v_{1}-v_{-1})}.
\end{equation}
We conclude from the second equation of (\ref{4d2}) that 
\begin{equation}\label{4d3}
\frac{f+L}{f-r}=\frac{v_1+L}{v_{1}-v_{-1}}K(v,\bar{v}_1).
\end{equation}
 Take the logarithmic derivative of (\ref{4d3}) with  respect to $v_{-1}$ and get $f-r=r_{v_{-1}}(v_1-v_{-1})$. 
Differentiation of the last equality with respect to $v_1$ yields  $f_{v_1}=r_{v_{-1}}$. 
We differentiate (\ref{4d2}) with respect to $v_{-1}$ and use the fact that  $f_{v_1}=r_{v_{-1}}$, we obtain 
$f_{v_1}=\pm\frac{f-r}{v_1-v_{-1}}$. 

First assume that $f_{v_1}=-\frac{f-r}{v_1-v_{-1}}$. We have, $f-r=D(v, v_{-1}, \bar{v}_{1})(v_1-v_{-1})^{-1}$. 
It follows from $r_{v_{-1}}=-\frac{f-r}{v_1-v_{-1}}$ that $f-r=C(v, v_1, \bar{v}_1)(v_{1}-v_{-1})^{-1}$, and, 
therefore, $f-r=C(v,  \bar{v}_1)(v_{1}-v_{-1})^{-1}$. We substitute this expression for $f-r$ into 
(\ref{4d3}) and see that $f+L=C(v, \bar{v}_1)K(v, \bar{v}_1)(v_1+L)(v_1-v_{-1})^{-2}$ which is impossible since $f$ 
does not depend on $v_{-1}$.

Now consider the case when $f_{v_1}=\frac{f-r}{v_1-v_{-1}}$. We have, 
$f-r=(v_1-v_{-1})D(v, v_{-1}, \bar{v}_{1})$. Also, $r_{v_{-1}}=\frac{f-r}{v_1-v_{-1}}$ implies that 
$f-r=(v_{1}-v_{-1})C(v, v_1, \bar{v}_1)$. One can see that $D(v, v_{-1}, \bar{v}_{1})=
C(v, v_1, \bar{v}_1)=:C(v, \bar{v}_1)$. Therefore, $f-r=C(v, \bar{v}_1)(v_1-v_{-1})$. It follows from 
(\ref{4d3}) that 
\begin{equation}\label{4d4}
f=A(v, \bar{v}_1)v_1+A(v, \bar{v}_1)L-L, 
\end{equation}
where $A=CK$. Note that $A=A(v, \bar{v}_1)$ and $A_1=A(v_1, f(v, v_1, \bar{v}_1))$. 
Substitute (\ref{4d4}) into (\ref{4d1}), get
$$
(Av_1+AL-L-\bar{v}_1)(A_1v_2+A_1L)(v_2-v)(v_1+L)
$$
$$
=(A_1v_2+A_1L-L-\bar{v}_1)(Av_1+AL)(v_1-v)(v_2+L),
$$
and compare the coefficients before $v_2^2$, we have
\begin{equation}\label{4d5}
A_1(Av_1+AL-L-\bar{v}_1)(v_1+L)=A_1(Av_1+A)(v_1-v).
\end{equation}
 It follows from (\ref{4d5}) that $A_1=0$ or, by comparing the coefficients before $v_1$, one
 gets $A=\frac{L+\bar{v}_1}{L+v}$. Therefore, by (\ref{4d4}), we have the equation $v_{1,1}=f=\frac{L(\bar{v}_1+v_1-v)+v_1\bar{v}_1}{L+v}$. Note that the equation is symmetric
 with respect to variables $v_1$ and $\bar{v}_1$. This observation allows one to write down 
an $n$-integral $I$ by a given $m$-integral $\bar{I}$ by changing in $\bar{I}$   
variables $v_j$ into variables $\bar{v}_j$, $j=1,2$.

\noindent \underline{Case $7^{**}$}: Consider all equations (\ref{discrete})  with $m$-integral
 $F=2v-v_1-v_{-1}=D^{-1}\bar{I}$, where $\bar{I}=2v_1-v-v_2$.
Equation $v_{1,1}=f(v,v_1,\bar{v}_1)$  can be rewritten as $v_{-1,1}=r(v,v_{-1},\bar{v}_1)$.
Equality $\bar{D}F=F$ implies 
\begin{equation}\label{7*}
2\bar{v_1}-f-r=2v-v_1-v_{-1}.
\end{equation}
We apply $\frac{\partial}{\partial{v_1}}$ and $\frac{\partial}{\partial{v_{-1}}}$ to (\ref{7*}) 
and find that $f_{v_1}=1$ and $r_{v_{-1}}=1$. Therefore, $f=v_1+h(v,\bar{v}_1)$ and 
$r=v_{-1}+q(v,\bar{v_1})$. Substitute these expressions for $f$ and $r$  into $(\ref{7*})$ and get
\begin{equation}\label{7**}
q=2\bar{v}_1-2v-h.
\end{equation}
Equation $v_{1,1}=f=v_1+h(v,v_1)$ can be rewritten as
\begin{equation}\label{7***}
\bar{v}_1=v+h(v_{-1},v_{-1,1})=v+h(v_{-1},v_{-1}+q(v,\bar{v}_1)).
\end{equation}
First differentiate (\ref{7***}) with respect to ${v_{-1}}$ and then apply the shift operator $D^{-1}$, we get
$D^{-1}h_v+D^{-1}h_{\bar{v}_1}=0$, that is  $h=h(\bar{v_1}-v)$. Equations $(\ref{7*})$ -  $(\ref{7***})$ 
give $\bar{v}_1-v=h(2\bar{v}_1-2v-h)$, or by taking $\epsilon=\bar{v_1}-v$ one gets 
$\epsilon=h(2{\epsilon}-h(\epsilon))$.
Therefore, the equation with $m$-integral $\bar{I}=2v_1-v-v_2$ becomes $v_{1,1}=v_1+h(\bar{v}_1-v)$, 
where $h$ solves a functional equation $\epsilon=h(2{\epsilon}-h(\epsilon))$.
This equation  $v_{1,1}=v_1+h(\bar{v}_1-v)$  admits also an $n$-integral. Since the equation is of the form $Dz=h(z)$ with $z=\bar{v}_1 -v_1$ then we have $D(z-h^{-1}(z))=z-h^{-1}(z)$. Actually, $D(z-h^{-1}(z))=D(z)-z=h(z)-z=z-h^{-1}(z)=z-h^{-1}(z)$. Here we use the identity $h(z)-z=z-h^{-1}(z)$ 
which is equivalent to the functional equation $z=h(2z-h(z))$.

\section{Conclusions}

The  problem of discretization of Liouville type equations is discussed. Besides  purely theoretical interest as a bridge between two parallel realizations of the integrability theory, this subject has an important practical significance. There are two-dimensional Toda field equations corresponding to each semisimple or of Kac-Moody type Lie algebra (see \cite{MOP}, \cite{DrinfeldSokolov}). The question is open whether there exist integrable
discrete versions of these. Different particular cases are studied in \cite{Suris}, \cite{Ward}, \cite{Habibullin}. In the article a step is done towards the solution of the problem. An effective method of discretization is suggested based on integrals. It is known that the B\"{a}cklund transform is a kind of discretization (see  \cite{AdlerStartsev}, \cite{ShabatVeselov}). We would like to stress that our method of discretization essentially differs from that one.  Even though for some exceptional cases the semi-discrete equation obtained realizes the B\"{a}cklund transformation of the original equation for the other examples it is not the case.

\section*{Acknowledgments}
This work is partially supported by the Scientific and
Technological Research Council of Turkey (T\"{U}B{\.{I}}TAK)  grant $\# $209 T 062,
Russian Foundation for Basic
Research (RFBR) (grants $\#$ 11-01-00732-a, $\#$
11-01-97005-r-povoljie-a, $\#$ 10-01-91222-CT-a and $\#$ 10-01-00088-a), and MK-8247.2010.1.

\end{document}